# Life Under Your Feet: An End-to-End Soil Ecology Sensor Network, Database, Web Server, and Analysis Service

Katalin Szlavecz[†], Andreas Terzis[*], Stuart Ozer[+],
Razvan Musăloiu-E.[*], Joshua Cogan[‡], Sam Small[*]
Randal Burns[*], Jim Gray[+], Alex Szalay[‡]

Computer Science Department[*], Department of Earth and Planetary Sciences[†], Department of Physics and Astronomy[‡]
The Johns Hopkins University
Microsoft Research[+]

**June 2006**
Microsoft Technical Report MSR TR 2006 90

**Abstract**[1]: Wireless sensor networks can revolutionize soil ecology by providing measurements at temporal and spatial granularities previously impossible. This paper presents a soil monitoring system we developed and deployed at an urban forest in Baltimore as a first step towards realizing this vision. Motes in this network measure and save soil moisture and temperature in situ every minute. Raw measurements are periodically retrieved by a sensor gateway and stored in a central database where calibrated versions are derived and stored. The measurement database is published through Web Services interfaces. In addition, analysis tools let scientists analyze current and historical data and help manage the sensor network. The article describes the system design, what we learned from the deployment, and initial results obtained from the sensors.

The system measures soil factors with unprecedented temporal precision. However, the deployment required device-level programming, sensor calibration across space and time, and cross-referencing measurements with external sources. The database, web server, and data analysis design required considerable innovation and expertise. So, the ratio of computer-scientists to ecologists was 3:1. Before sensor networks can fulfill their potential as instruments that can be easily deployed by scientists, these technical problems must be addressed so that the ratio is one nerd per ten ecologists.

## 1. Introduction

Lack of field measurements, collected over long periods and at biologically significant spatial granularity, hinders scientific understanding of how environmental conditions affect soil ecosystems. Wireless Sensor Networks promise a fountain of measurements from low-cost wireless sensors deployed with minimal disturbance to the monitored site.

In 2005 we built and deployed a soil ecology sensor network at an urban forest. The system, called *Life Under Your Feet*, includes:

*Motes* are embedded computers that collect environmental parameters such as soil moisture and temperature and periodically send their measurements to gateways.

*Gateways* are static and mobile computers that receive status updates from motes and periodically download collected measurements to a database server.

*Database* stores measurements collected by the gateways, computes derived data, and performs data analysis tasks.

*Calibration algorithms* convert raw measurements into scientific values like temperature, dew point, etc, that are stored in the database

*Access and analysis tools* allow us to analyze and visualize the data reported by the sensors.

*Web site* serves the data and tools to the Internet.

*Monitors* are programs that observe all the aspects of the system and generate alerts when anomalies occur.

The unique aspects of *Life Under Your Feet* are: (1) Unlike previous wireless sensor networks *all* the measurements are saved on each mote's local flash memory and periodically retrieved using a reliable transfer protocol. (2) Sophisticated calibration techniques translate raw sensor measurements to high quality scientific data. (3) The database and sensor network are accessible via the Internet, providing access to the collected data through graphical and Web Services interfaces.

This system is only a first step in the arduous process of transforming raw measurements into scientifically important results. However, it promises to improve ecology and ecologists' productivity – and we believe it has implications for other disciplines that collect sensor data. Today the project has one ecologist and several supporting computer scientists. We are working to reverse that ratio.

The rest of the paper is structured as follows: Section 2 provides background information on soil ecology, how sensor networks can help gather data from field deployments, and the requirements for doing so. Sections 3 and 4 present the data collection and publishing system design. Section 5 presents results from a six-month deployment, and Section 6 we presents the lessons we learned from this deployment. Section 7 summarizes the paper and suggests future research directions.

---

[1] An earlier (and shorter) version of this article appeared in [Musăloiu-E 2006].

## 2. Soil Ecology

Soil is the most spatially complex stratum of a terrestrial ecosystem. Soil harbors an enormous variety of plants, microorganisms, invertebrates and vertebrates. These organisms are not passive inhabitants; their movement and feeding activities significantly influence soil's physical and chemical properties. The soil biota are active agents of soil formation in the short and long term. At the same time, soil is an important water reservoir in terrestrial ecosystems and, thus, an important component for hydrology models. All these factors play fundamental roles in Earth's life support system. But, we poorly understand their interactions because of the enormous diversity of these organisms, and the complex ways they interact with their environment [Wardle2004], [Young2004].

Among the major challenges in studying soil biota are the size range (from micrometers to centimeters,) their diversity, their sparse yet-clustered population distribution, and the enormous spatial and temporal heterogeneity of the soil substrate.

Soil organism population densities are skewed in all three dimensions. Often these distributions reflect diversity of the physical environment, because many soil invertebrates are sensitive to such abiotic factors as soil moisture, temperature, and light. Most species are negatively phototactic, i.e. tend to move away from light, although the diurnal cycle is still important in determining animal activity. Population aggregations can be biologically driven i.e. animals are 'attracted' to each other [Takeda1980], or they create favorable microhabitats for one another [Szlavecz1985]. More frequently, patches of favorable abiotic conditions or resources are the underlying cause, but sometimes there is no obvious physical or biological mechanism behind these aggregations [Jimenez2001].

It is important to emphasize, that soil organisms are not just passively reacting to abiotic conditions; rather, they are active factors of soil formation influencing many of its physical, chemical and biological properties. Earthworms are often called ecosystem engineers or keystone organisms, because of their major role in soil processes. By feeding on detritus and mixing organic and mineral layers the profoundly affect soil aggregate stability, pore size distribution, carbon storage and turnover and thus indirectly plant growth. All these changes ultimately affect soil water holding capacity, therefore soil moisture conditions, which is a major abiotic factor determining earthworm distribution and abundance.

Any field study of soil biota includes information on weather, soil temperature, moisture, and other physical factors. These data are usually collected by a technician visiting the field site once a week, month, or season and taking a few spatial measurements that are subsequently averaged. Therefore, only a few measurements per site are available. These techniques are labor-intensive and do not capture spatial and temporal variation at scales meaningful to understand the dynamics of for soil biota. Moreover, frequent visits to a site disturb the habitat and may distort the results. Some sites are not easily accessible, e.g. monitoring wetland soils can be challenging, and some site visits involve property issues.

The ecologist in the team works with the Baltimore Ecosystem Study LTER ([www.beslter.org](www.beslter.org)). The project focuses on urban ecosystems, and much of the field sampling takes place in residential areas. So far homeowners have been exceptionally cooperative and supportive to our work. A small device deployed on their property and taking environmental measurements is much less intrusive than a field technician trampling through their yards on a regular basis.

Clearly, using in-situ sensors that can report results continuously and without visiting the site would be a huge productivity gain for ecologists. Such sensors could give them more data without perturbing the site after the installation. But, until recently, continuous-monitoring data loggers were prohibitively expensive. That is about to change.

### 2.1. Requirements

Sensor systems promise inexpensive, hands-free, low-cost and low-impact ecological data collection — an attractive alternative to manual data logging — in addition to providing considerably more data at finer spatial and temporal granularity. However, to be of scientific value, the data collection design should be driven by the experiment's requirements, rather than by technology limitations. Here are the key requirements for soil ecology sensor systems:

**Measurement Fidelity:** *All* the raw measurements should be collected and persistently stored. Should a scientist later decide to analyze the data in different ways, to compare it to another dataset, or to look for discrepancies and outliers, the original data must be available. Furthermore, given the communal nature of field measurement locations, other scientists might use the data in ways unforeseen when the original measurements were taken. Generally speaking, *techniques that distill measurements for a specific purpose potentially discard data that are important for future studies*. Both the raw and distilled data should be preserved.

**Measurement Accuracy and Precision:** Research objectives should drive the desired accuracy. For example, while temperature variation of half a degree does not directly affect soil animal activity, soil respiration increases exponentially with temperature, so half a degree makes a significant difference. Movement and storage of soil water is another good example. Most soil moisture sensors estimate soil water using a calibrated relationship between moisture content and another measurable variable (*e.g.* dielectric constant, electrical resistance). Measurement output can be volumetric moisture content or water potential. Choice of technique and desired accuracy depends on the project goal (in addition to the obvious factors such as cost, duration of the experiment, etc). Calculating evapotranspiration rates for plant-soil interaction research requires more accurate measurements than deciding when to irrigate. Plant physiology studies and hydrology models need data on water pressure, while most soil invertebrate studies are interested in volumetric water content. In the latter case 1% change may not affect activity as long as it is within the species' optimal range. However, if moisture content approaches the upper and lower species' tolerance limits, even small changes may have big effects in activity or even survival. Again, soil respiration and in general, soil microbial activity is a function of soil moisture. Therefore, *raw measurements need to be precisely calibrated, to give scientists high confidence that measured variations reflect changes in the underlying processes rather than random noise, systematic errors, or drift.*

**Sampling Frequency:** While fixed sampling periods are adequate for most tasks, there are scenarios where variable sampling rates are desirable. Hourly sampling is adequate for most environmental monitoring; however, during an extreme event such as a rainstorm, one wants to sample more frequently (*e.g.* every 10 minutes). In other cases – sampling gas concentrations, for example – preliminary measurements are necessary to determine the optimal sampling frequency. It is evident from the above that *the system should support a dynamic sampling frequency, at minimum based on external commands and potentially based on application-aware logic implemented in the network.*

**Fusion with External Sources**: *Comparing measurements with external data sources is crucial*. For instance, soil moisture and temperature measurements must be correlated with air temperature, humidity, and precipitation data. Animal activity is determined by these factors as much as by soil temperature and moisture. In the case of hydrology models, one can only make sense of soil moisture if precipitation data is available. In addition to "traditional" external data sources such as weather stations, data from other sensor systems can be useful. Hence, the sensor net, should export it data using a controlled vocabulary and well defined schema and formats.

**Experiment Duration**: Some ecological studies, such as identifying the interactions between plant growth and soil water, require measurements on short temporal scales ─ a single growing season or a few years. But, *measurements for ecosystem studies generally last several years*. This makes per-mote battery-powered deployments infeasible. In these cases, alternative energy sources such as energy harvesting are necessary [Jiang2005]. The scientific questions underlying the deployment drive the experiment's duration. At one extreme, scientists might want to observe long-term changes: How do soil conditions change during secondary succession after clear cutting? Such an experiment would last at least fifty years. The primary goal of the he NSF-funded Long Term Ecological Research (LTER) System is to investigate ecological processes over long temporal and broad spatial scales (http://www.lternet.edu/). Such long-term monitoring has become essential to provide data on climate change and other global environmental issues (e.g. melting of permafrost and subsequent carbon release, altered soil conditions in urban environments, effect of no-till farming on soil moisture, etc).

**Deployment Size:** Scientists have very little information about the size of underground organism population-patches. Therefore, the spatial measurement requirements are not known. This is typical of the current state of ecological measurement. For example, to observe earthworm aggregations one needs at least a 10 x 5 grid with the grid-points 5-10 m apart – but a finer grid would be better. In many cases, using a grid is not the preferred sampling method. For instance, scientists would like to deploy ecology sensor systems in lawns, flowerbeds, vegetable gardens, and other land cover types. In these cases, the emphasis is on the land cover categories, as they presumably drive population skew. Therefore, *systems should be deployed in ways that capture the heterogeneity of land use on multiple scales.*

## 3. System Architecture

Figure 1 depicts the overall architecture of the system we developed and deployed during the Fall of 2005 in an urban forest adjacent to the Homewood campus of the Johns Hopkins University. Each of the deployed motes measures soil conditions. The collected measurements are stored on the motes' local flash memory and are periodically retrieved by a sensor gateway over a single-hop wireless link. The raw measurements retrieved by the gateway are inserted into a SQL database. They are then calibrated using sensor-specific calibration tables and are cross-correlated with data from external data sources (*e.g.* data from the weather service and from other sensors). The database acts both as a repository for collected data and also drives data conversion. Data analysis and visualization tools use the database and provide access to the data through SQL-query and Web Services interfaces.

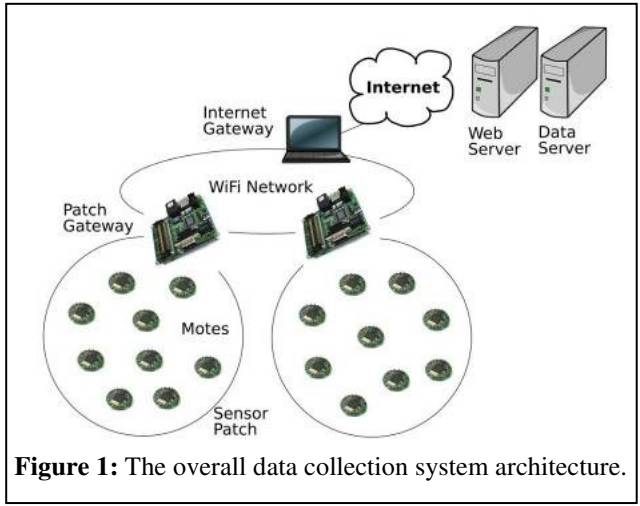

**Figure 1:** The overall data collection system architecture.

### 3.1. Motes and Sensors

A mote platform that meets the requirements outlined in Section 2.1 must be relatively low-cost, energy-efficient, user-programmable (to collect data from custom sensors), and have wireless communication capabilities. With these objectives in mind, we selected the popular MICAz mote from Crossbow [Crossbow], [MICAz].

MICAz is a user-programmable device using a Atmel ATMEGA 128L microcontroller with 128 KB of program memory and 4 KB of RAM, 7 Analog to Digital converters (ADC) with 10-bit resolution, and 512 KB of flash for persistent storage. It also has a CC2420 802.15.4 radio transceiver capable of 250Kbps at 100 m range [TI]. Each MICAz has a Crossbow MTS101 data acquisition board [MTS] for custom sensor interfaces. The MTS101 includes an ambient light and temperature sensor in addition to connections for 5 external sensors. We designed a custom waterproof case for the whole assembly powered by two AA batteries (Figure 2.)

The MICAz motes run software we developed on TinyOS, an open-source operating system for wireless embedded sensor systems [Hill2000]. Using component libraries from TinyOS and our own written using nesC [Gay2003], we are able to customize the motes to support our sensors, meet our deployment requirements, and control its behavior.

The TMote Sky mote [MotIV] also meets our requirements. Its capabilities are comparable to the MICAz, but has lower power consumption in most operating modes, is equipped with integrated light, temperature, and humidity sensors, and is directly programmable via an on-board USB connector (an external programming board is required for MICAz motes). The TMote has 12-bit ADCs compared to the 10 bits of resolution provided by MICAz. On the other hand, a significant benefit of MICAz is its 51-pin expansion connector. This allowed us to design, prototype, and test our custom sensors without direct soldering to the mote via the MTS101 data acquisition board. The deciding factor was ultimately the flexibility of the MICAz platform compared to the longer lifetime offered by TMote.

### 3.2. Sensor Interfaces and Drivers

The motes are equipped with Watermark soil moisture sensors, which vary resistance with soil moisture, and soil thermistors which vary resistance with temperature. Watermark soil moisture sensor respond well to soil wetting-drying cycles following rain events [Shock200], and are inexpensive —an important issue for large deployments. Both sensor types were purchased from Irrometer [Irrometer].

These sensors report changes in physical parameters by changing their resistance. Since the analog to digital converter digitizes voltage readings, we built a voltage divider that varies the ADC voltage as the sensor resistance changes by connecting a 10 kΩ resistor between power and the ADC pin and connecting the sensor to the ADC pin and ground. This uses a power pin and an ADC pin per sensor but eliminates the need for a multiplexer.

The TinyOS device driver we developed for the moisture and temperature sensors are similar to the ones used for the photo and temperature sensors on the MTS101.

### 3.3 Sensor Calibration

Knowing and decreasing the sensor uncertainty requires a thorough calibration process before deployment ― testing both precision and accuracy.

An evaluation the soil thermistors showed they are relatively precise (±0.5ºC), yet consistently returned values 1.5ºC below a NIST approved thermocouple. The 1.5ºC bias does not present a problem because we convert resistance to temperature using the manufacturer's regression technique. Furthermore, a 10 kΩ reference resistance is connected in series with the moisture sensors on each mote. Since the resistance's value directly factors into the estimation of the sensor resistance, the bias is measured individually, recorded in the database, and used during the conversion from raw to derived temperature.

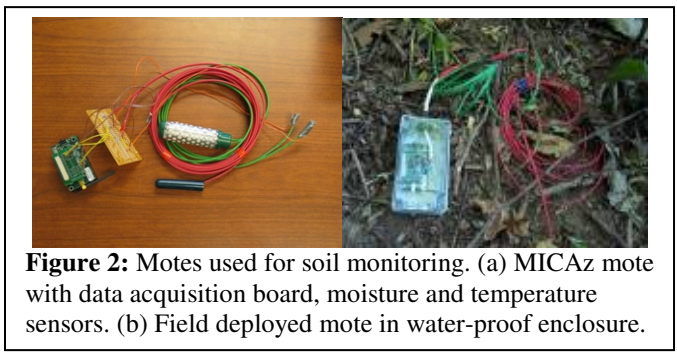

**Figure 2:** Motes used for soil monitoring. (a) MICAz mote with data acquisition board, moisture and temperature sensors. (b) Field deployed mote in water-proof enclosure.

The temperature sensors are easily calibrated; their output is a simple function of temperature. However, each moisture sensor requires a unique two-dimensional calibration function that relates resistance to both soil moisture and temperature. Each moisture sensor is calibrated individually by measuring resistance at nine points (three moisture contents each at three temperatures) and using these values to calculate individual coefficients to a published regression [Shock1998]. Moisture sensor precision was tested with eight sensors in buckets of wet sand measuring their resistance every ten minutes, while varying the temperature from 0ºC to 35ºC over 24 hours. We found that six sensors gave similar readings, but two did not.

## 3. 4. Data Collection Subsystem

We programmed the motes to sample each onboard sensor once a minute and store the data in a circular buffer in their local flash. Using flash memory allows retrieving all observed data over lossy wireless links — in contrast to *sample-and-collect* schemes such as TinyDB which can lose up to 50% of the collected measurements [Tolle2005]. Since each mote collects approximately 23 KB per day, the MicaZ 512 KB flash can buffer for 22 days. In practice, sensor measurements are downloaded from the motes weekly or at least once every two weeks. To allow on-line monitoring, each mote periodically broadcasts a series of status messages. During the testing period, these broadcasts happen every two minutes – but to extend battery life, the broadcasts could be once an hour. Each status message contains the mote's ID, the amount of data currently stored, the current battery voltage reading, and a link-quality indicator (LQI)[2]. The message exchanges during the status report phase are depicted in Figure 3 (a). Immediately after turning the radio on, the mote sends a status message to signal its presence. During the 2 seconds that the radio is active, the mote sends 5 more status messages, each 250 milliseconds apart. The mote turns its radio off until the next status report to conserve energy, if the base does not make any requests during this period,.

The base station periodically retrieves collected samples from each of the motes in the network as shown in Figure 3.b. Upon receiving a status message from the mote, the base may issue a download request for all new data since a specified time. This *Bulk Phase* concludes with the mote transmitting another status message. Radio packets may be lost due to the variable radio link quality. The base station maintains a list of "holes" signifying missing or malformed (*e.g.,* bad CRC) packets. A NACK-based automatic repeat request (ARQ) protocol recovers these lost packets during the *Send-and-Wait Phase* in which the base station sequentially requests each missing data packet. This phase concludes when all the missing data segments have been recovered.

## 4. Database Design

The database design (Figure 4) follows naturally from the experiment design and the sensor system. Each entry in the `Site` table describes a geographic region with a distinct character (e.g. an urban woodland or a wetland). All the sites in our case are in the Greater Baltimore area, for which common macro-weather patterns apply. Each site is partitioned into `Patches`. Each patch is a coherent deployment area, defined through its GPS coordinates. Each patch contains `Motes`. A particular mote has an array of `Sensors` that report environmental measurements. Mote and sensor locations are precisely located relative to the reference coordinates of a patch.

The Mote and Sensor types (metadata) are described in corresponding `Type` tables. Each mote has a record in the `Motes` table describing its model, deployment, and other metadata. Each `Sensor` table entry describes its type, position, calibration information, and error characteristics. The `Event` table records state changes of the experiment such as battery changes, maintenance, site visits, replacement of a sensor, sensor failure, etc. Global events are represented by pointing to the NULL patch or NULL Mote. The site configuration tables (`Site`, `Patch`, `SiteMap`) hardware configuration tables (`Mote`, `Sensor`, `MoteType`, `SensorType`), and sensor calibrations (`DataConstants`, `RToSoilTemp`) are loaded prior to data collection. As new motes or sensors are added, new records are added to those tables. When new types of mote or sensor are added, those types are added to the type tables.

Measurements are recorded in the `Measurement` table which has a timestamped entry containing each raw value reported by a mote. The `Measurement` table is actually a "wide" vector-of values today because all the motes report the same data; but the table should be pivoted *(sensor,time,value)* to support a more heterogeneous sensor system in the future. Figure 1 shows that pivoted schema. Calibrated versions of the data and derived values are recorded in the `Calibrated` table. External weather data is recorded in the `WeatherInfo` table. Various support tables contain lookup values used in sensor calibration.

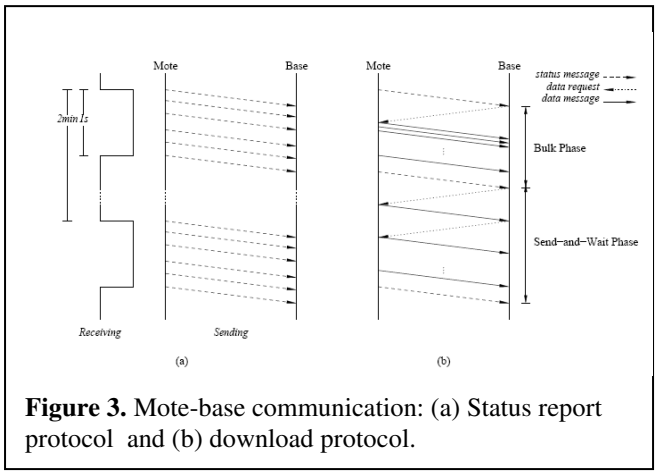

**Figure 3.** Mote-base communication: (a) Status report protocol and (b) download protocol.

---

[2] The LQI is provided by the mote connected to the base-station that receives the status report.

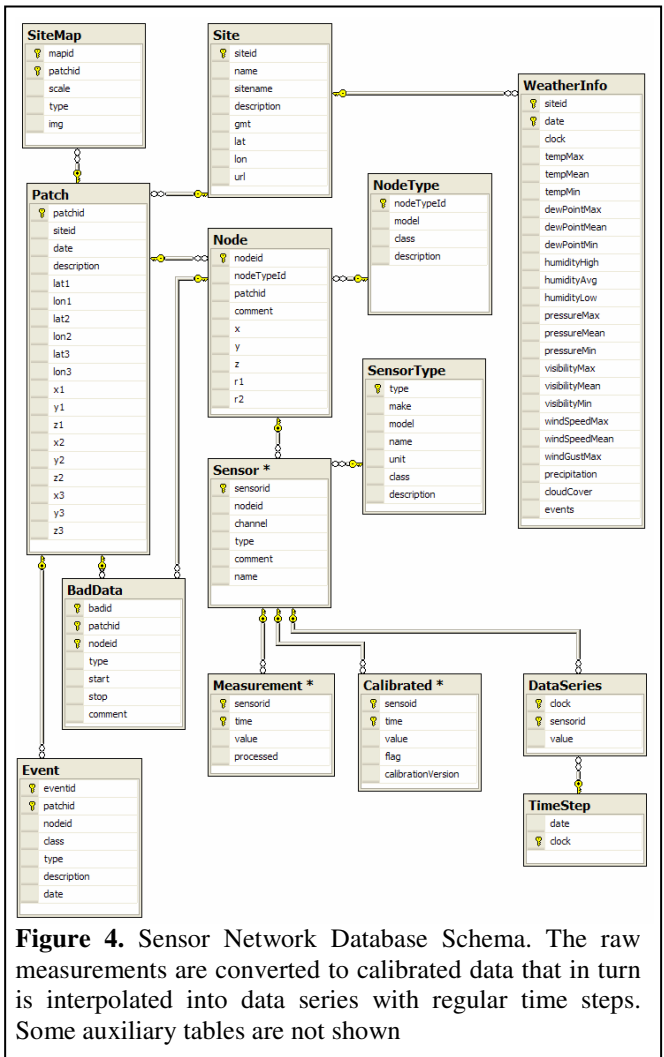

**Figure 4.** Sensor Network Database Schema. The raw measurements are converted to calibrated data that in turn is interpolated into data series with regular time steps. Some auxiliary tables are not shown

The database, implemented in Microsoft SQL Server 2005, benefits from the skyserver.sdss.org database and website design and support procedures built for Astronomy applications [SDSS]. The new website inherited the SkyServer's self-documenting framework that uses embedded markup tags in the comments of the data definition scripts to characterize the metadata (units, descriptions, enumerations, for the database objects, tables, views, stored procedures, and columns.) The data definition scripts are parsed to extract the metadata information and insert it into the database. A set of stored procedures generate an HTML rendering of the hyperlinked documentation (see the `Schema-Browser` tab on [LifeUnderYourFeet]).

### 4.1. Loading Raw Data

The initial deployment collected 1.6M mote readings (soil moisture, soil temperature, ambient temperature, ambient light, and battery voltage), for a total of 6M measurements. Raw measurements arrive from the gateway as comma-separated-list ASCII files. The loader performs the two-step process common to data warehouse applications. (1) The data are first loaded into a quality-control (QC) table in which duplicate records and other erroneous data are removed. (2) Next, the quality-controlled data are copied into the `Measurement` table, with the `processed` flag set to 0.

In the terminology of NASA's *Committee on Data Management, Archiving, and Computing (CODMAC) Data Level Definitions* [CODMAC], this input data is *Level 0* data (raw time-space data) that is transformed to *Level 1* data by converting "sensor time" to GMT, and by geo-locating the measurements. These transformations are invertible and lossless, so the Level 0 data can be reconstructed from the Level 1 data. Consequently, once the Level 0 data is moved to the Level 1 `Measurement` table, the contents of the QC table are purged.

### 4.2 Deriving Calibrated Measurements

The raw data is converted to scientifically meaningful values by a multistage program pipeline run within the database as SQL stored procedures. These procedures are triggered by timers or by the arrival of new data. The conversions apply to all `Measurement` values with `processed = 0`. Each conversion produces a calibrated measurement for the `Measured` table, and sets the value's `Measurement.processed = 1`.

As explained in Section 3.3, the raw sensor data voltages are converted to science data using sensor-specific algorithms that often need other environmental data. The conversion takes an unprocessed "row" from the `Measurement` table and computes several derived values.

As shown in Figure 5, calibrated data is saved in the `Calibrated` table, where each measurement from each sensor is stored in a separate row (*i.e.,* the data is pivoted on (time, sensor, value, StdError)).

The calibrated data is aggregated and gridded into the `DataSeries` table, which contains calibrated data values averaged over a predefined intervals, defined by the `TimeStep` table. This time-and-space gridded `DataSeries` representation is convenient for analysis.

In the *CODMAC Data Level Definitions* [CODMAC], this is a `conversion from` *Level 1* data (raw time-space data) to *Level 2* `Measures` data (calibrated science data), and the averaged, interpolated, and time-gridded `DataSeries` data is *Level* 3 data.

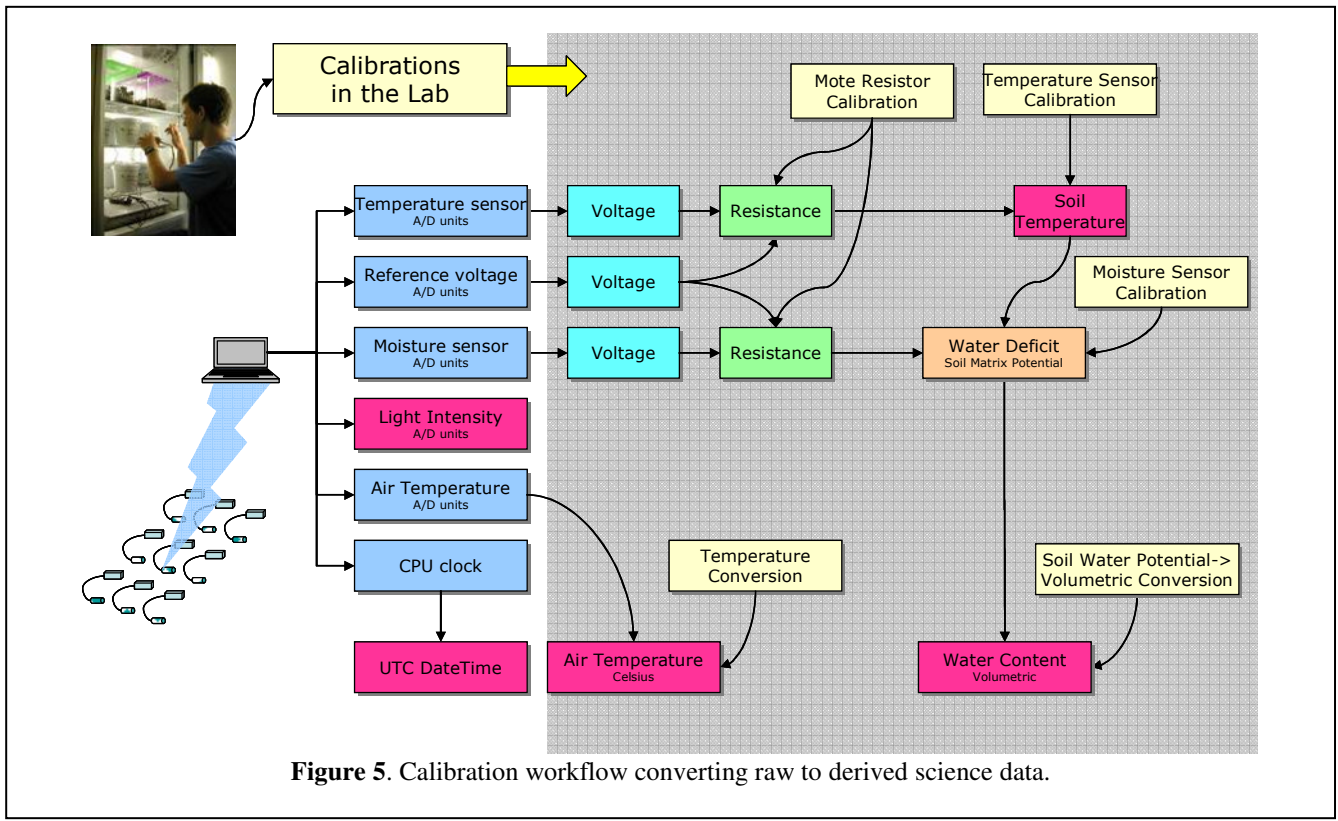

**Figure 5**. Calibration workflow converting raw to derived science data.

Each load and calibration step is recorded in the `LoadHistory` table, with the input filename, the timestamp of the loading, and its own unique `loadVersion` value, and some metadata information about what procedures were used, and what errors were seen. This `LoadVersion` value is also saved with every entry in the `Measurement` table and the version of the calibration software is recorded in each `Calibrated` table entry. This tracks data provenance (*i.e.,* the origin of each data value).

Figure 5 illustrates the data flow in the calibration pipeline that provides the precision and accuracy necessary for sensor-based science. Since soil moisture sensors have strong temperature dependence, an average soil temperature at each time step is used to calibrate moisture measurements for motes without a soil temperature value. This allows meaningful moisture results for all sensors.

We are currently implementing a database representation of the calibration workflow, representing the workflow as a graph, with the processing steps connecting the motes.

Some calibrated data is known to be bad. These intervals are represented in a `BadData` table, and the corresponding rows in the `Measurement` table are marked with an `isBad=1` flag, and these data values are never copied into the `Calibrated` table. For example, the interface boards on some sensors had loose connections for a while. As a result, some these measurements were invalid. Those intervals are represented in the `BadData` table.

There are two ways to deal with missing data, either interpolate over them, or treat them as missing. We believe that both approaches are necessary, their applicability depends on the scientific context. In any case, in the database the processing history must be clearly recorded, so that we can always tell how the calibrated data was derived from the raw measurements.

Background weather data from the Baltimore (BWI) airport is harvested from [wunderground.com](wunderground.com) and loaded into the `WeatherInfo` table. This data includes temperature, precipitation, humidity, pressure as well as weather events (rain, snow, thunderstorms, etc). In the next version of the database the weather data will be treated as values from just other sensors.

## *4.3. Web Data Access*

The current and historical sensor data and measurements are available from the website via standard reports. These reports present the data in tabular and graphical form with at common aggregation levels. The reports are useful for doing science and are also useful for managing the sensor system.

The reports present tabulated values for all the sensors on a given mote or for one sensor type across all motes (see http://lifeunderyourfeet.org/en/tools/visual/timeseries.aspx.) Another display shows the motes on a map with the sensor values modulating the color (see http://lifeunderyourfeet.org/SensorMap/MapView.aspx.)

The time series data can also be displayed in a graphical format, using a .NET Web service. The Web service generates an image of the raw or calibrated data series with the option to overlay the background weather information: temperature, humidity, rainfall, etc.

The web user interface and reporting tools need considerably more work -- soil scientists do not want to learn SQL and they often want to see graphical and spatial displays rather than tables of numbers.

They often want to see the aggregated sensor responses to discrete events like storms, cold-fronts or heat waves. For example: how does soil moisture vary as a function of time after a rain? We plan to provide spatial and temporal interpolation tools that answer questions such as: what is the soil moisture at the position of a sample of soil animals collected at a given time, from a certain depth? Eventually we will need to cross-correlate these interpolated values with results from other experiments.

As a stop-gap, and as a way to allow arbitrary analysis, the web and web-service interfaces expose the SQL Schema and allow SQL queries directly to the database: http://lifeunderyourfeet.org/en/help/browser/browser.asp and http://lifeunderyourfeet.org/en/tools/search/sql.asp. This guru-interface has proven invaluable for scientists using the Sloan Digital Sky Survey [SDSS], and has already been very useful to us. If there is some question you want to ask that is not built-in, this interface lets you ask that question. In addition, we expect to implement the MyDatabase and batch job submission system similar to the CasJobs system implemented by the SkyServer [O'Mullane2005].

## *4.4. OLAP Cube for Data Analysis*

In addition to examining individual measurements and looking for unusual cases, ecologists want a high level view of the measured quantities; they want to analyze aggregations and functions of the sensor data and cross-correlate them with other biological measurements.

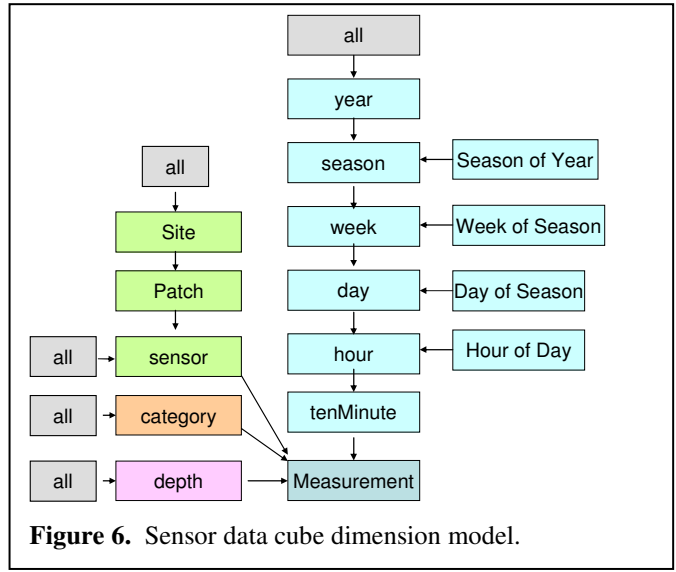

**Figure 6.** Sensor data cube dimension model.

The data is being collected to answer fundamental soil-science questions exploring both the time and spatial dimensions for small soil ecosystems. Typical questions we expect to answer are:

1. Display the temperature (average, min, max, standard deviation) for a particular time (*e.g.,* when animal samples are taken) or time interval, for one sensor, for a patch, for all sensors at a site, or for all sites. Show the results as a function of depth, time, as well as a function of patch category (land cover, age of vegetation, crop management type, upslope, downslope, etc).

2. Look for unusual patterns and outliers such as a mote behaving differently or an unusual spike in measurements.

3. Look for extreme events, *e.g.* rainstorms or people watering their lawns, and show data in time-after-event coordinates.

4. Correlate measurements with external datasets (*e.g.,* with weather data, the $CO_2$ flux tower data, or runoff data).

5. Notify the user in real-time if the data has unexpected values, indicating that sensors might be damaged and need to be checked or replaced.

6. Visualize the habitat heterogeneity, preferentially in three dimensions integrated with maps (*e.g.* LIDAR maps, with vegetation data, animal density data).

Queries 2-5 are standard relational database queries that fit the schema in Figure 4 very nicely —indeed the database was designed for them. But, Query 1 is really the main application of the data analysis and calls for a specialized database design typical of online analytical processing —a Data Cube that supports rollup and drill down across many dimensions [Gray1996]. The datacube and unified dimension model based on the relational database shown in Figure 6 follows fairly directly from the relational database design in Figure 4 It is built and maintained using the Business Intelligence Development Studio and OLAP features of SQL Server 2005.

The cube provides access to all sensor measurements including air and soil temperature, soil water pressure and light flux averaged over 10-minute measurement intervals, in addition to daily averages, minima and maxima of weather data including precipitation, cloud cover and wind.

The cube also defines calculations of average, min, max, median and standard deviation that can be applied to any type of sensor measurement over any selected spatio-temporal range. Analysis tools querying the cube can display these aggregates easily and quickly, as well as apply richer computations such as correlations that are supported by the multidimensional query language MDX [MDX]. Users can aggregate and pivot on a variety of attributes: position on the hillside, depth in the soil, under the shade vs. in the open, etc.

The cube aggregates the DataSeries fact table around three dimensions (when, who, where) – Time (`DateTimes`), Location/Sensor (`Sensor`), and Measurement Type (`MeasurementType`) (see Figure 6.)

The Time dimension includes a hierarchy providing natural aggregation levels for measurement data at the resolution of year, season, week, day, hour and minute (to the grain of 10-minute interval). Not only can data be summarized to any of these levels (*e.g.* average temperature by week), but this summarized data can then also be easily grouped by recurring cyclic attributes such as hour-of-day and week-of-year.

The Location/Sensor dimension includes a geographic hierarchy permitting aggregation or slicing by site, patch, mote or individual sensor, as well as a variety of positional or device-specific attributes (patch coordinates, mote position, sensor manufacturer, etc.) This dimension itself is constructed by joining the relational database tables representing sensor, site, patch and mote.

The weather data available in the cube uses these dimensions as well, although at a different time and space grain. In the Location/Sensor and time dimensions, weather is available per-site and per-day respectively. By sharing the same dimensions as the sensor measurements, relationships between weather and measurement information can be readily analyzed and visualized side-by-side using the tools.

Data visualization, trending and correlation analysis is most effective when measurement data is available for every 10-minute measurement interval of a sensor. While it is straightforward to handle large contiguous data gaps by eliminating a gap period from consideration, frequent gaps can interfere with calculations of daily or hourly averages. To avoid these problems, we plan to use interpolation techniques to fill any holes in the data prior to populating the cubes.

This OLAP data cube, using SQL Server Analysis Services, will be accessible via the Web and Web Services interface. We are experimenting with SQL Servers' built-in reporting services [Reporting Services], as well as the Proclarity [Proclarity], and Tableau [Tableau] data analysis tools that provide a graphical browsing interface to data cubes and interactive graphing and analysis.

## 5. Results

We deployed 10 motes into an urban forest environment nearby an academic building on the edge of the Homewood campus at Johns Hopkins University in September 2005. As Figure 7 illustrates, the motes are configured as a slanted grid with motes approximately 2m apart. A small stream runs through the middle of the grid; its depth depends on recent rain events. The motes are positioned along the landscape gradient and above the stream so that no mote is submerged.

A wireless base station connected to a PC with Internet access resides in an office window facing the deployment. Originally this base station was expected to directly collect samples from the motes. Once the motes were deployed, however, we discovered that some motes could not reliably and consistently reach the base station. Our temporary solution to this problem was to periodically visit the perimeter of the deployment site and collect the measurements using a laptop connected to a mote acting as base station.

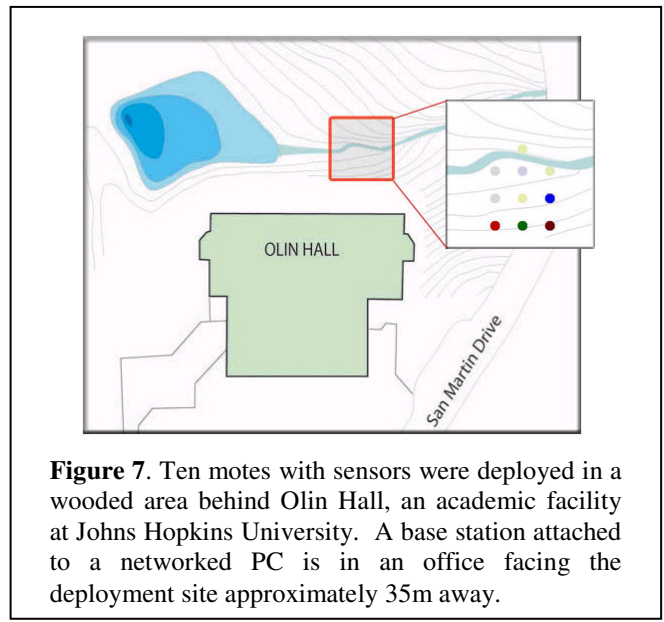

**Figure 7**. Ten motes with sensors were deployed in a wooded area behind Olin Hall, an academic facility at Johns Hopkins University. A base station attached to a networked PC is in an office facing the deployment site approximately 35m away.

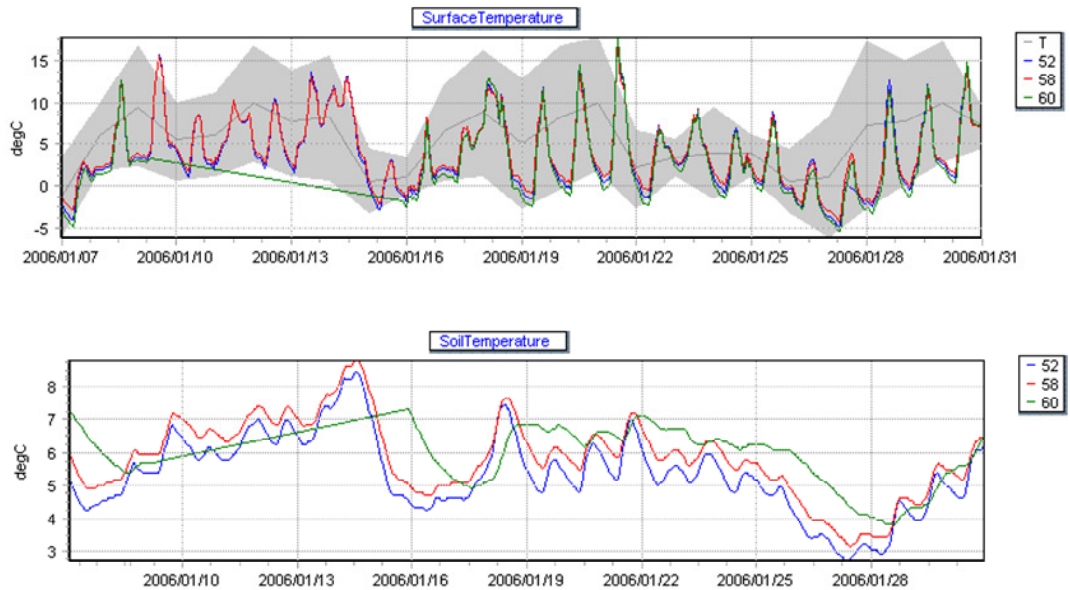

**Figure 8**. Air temperature data recorded by three motes at soil surface (upper figure) and at 10 cm depth (lower figure) during January 2006 (note the difference in the temperature scales. Data Shaded area is minimum and maximum air temperature for the Baltimore Metropolitan Area.

## *5.1. Ecology Results and Outlook*

During a 147 day deployment, the sensors collected over 6M data points. A subset of the temperature and moisture data is shown on Figures 8 and 9 respectively. Temperature changes in the study site are in good agreement with the regional trend. An interesting comparison can be made between air temperature at the soil surface and soil temperature at 10cm depth. While surface temperature dropped below 0ºC several times, the soil itself was never frozen. This might be due to the vicinity of the stream, the insulating effect of the occasional snow cover, and heat generated by soil metabolic processes. Several soil invertebrate species are still active even a few degrees above 0ºC and, thus, this information is helpful for the soil zoologist in designing a field sampling strategy.

Precipitation events triggered several cycles of quick wetting and slower drying. In the initial installation, saturated Watermark sensors were placed in the soil and the gaps were filled with slurry. We found that about a week was necessary for the sensor to equilibrate with its surrounding. Although the curves on Figure 9 reflect typical wetting and drying cycles, they are unique to our field site because the soil water characteristic response depends on soil type, primarily on texture and organic matter content [Munoz-Carpena2004].

We deliberately placed the motes on a slope, and our data reflect the existing moisture gradient. For instance mote 51 placed high on the slope showed greater fluctuations then motes 56 and 58, which were closer to the stream (see Figure 9). We occasionally performed synoptic measurements with *Dynamax Thetaprobe* sensors to verify our results.

Four of our current research topics within the Baltimore Ecosystem Study will benefit from the data provided by the sensor system:

1. **How do non-native become established and spread in urban areas?** Urban areas are "hotspots" for species introduction. The nature and extent of soil invertebrate invasions and the key physical and biological factors governing successful establishment are poorly known. [Johnston2003, 2004] Our hypothesis is that exotic species survive better in cities because they are less fluctuating environments. Population data show that both earthworm biomass and density are 2-3 times larger in urban forests [Szlavecz2006]. The sensor system will provide important data to two questions related to this topic: (1) Do urban and rural soil abiotic conditions in the same type of habitat differ? (2) Which elements of the urban landscape act as refuges for soil organisms during unfavorable periods? For instance irrigation of lawns and flowerbeds maintains a higher moisture level. In winter, the organisms can congregate around houses, or compost heaps, where the temperature is locally higher. Both examples promote both survival and longer periods of activity, which may result in greater number of offspring.

2. **What are the reproductive strategies of invasive species?** Although the exact mechanisms leading to successful invasion are poorly understood, the species' reproductive biology is often a key element in this process. In temperate regions, reproduction is closely tied to seasonal temperature changes. For terrestrial isopods the situation is more complicated, because hormonal changes necessary to initiate reproduction are also influenced by in light intensity and wavelength composition [Juchault1981], [Jassem1981]. Sensor systems can measure detailed temperature, light, and spectral flux both at the soil surface level, and at the strata within the soil where the organisms live.

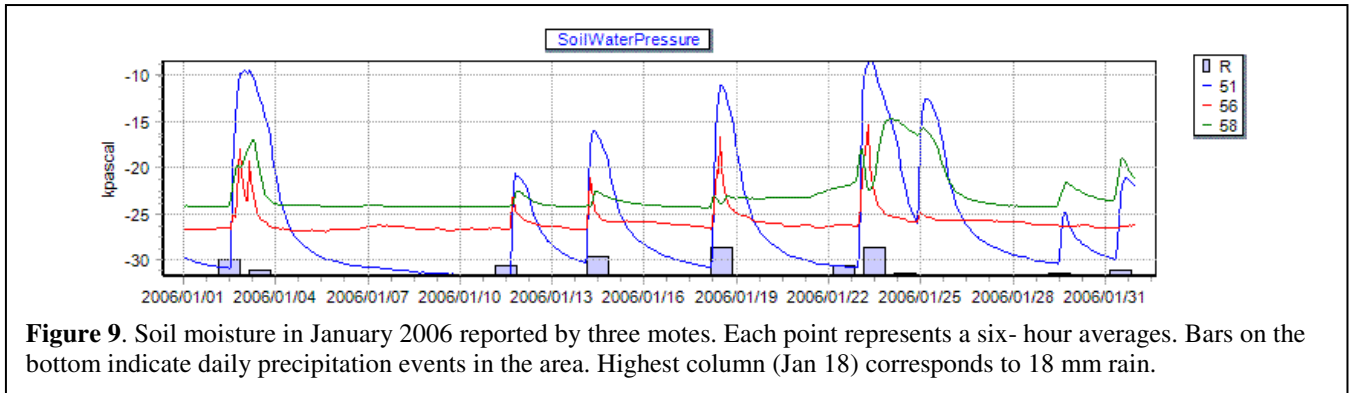

**Figure 9**. Soil moisture in January 2006 reported by three motes. Each point represents a six- hour averages. Bars on the bottom indicate daily precipitation events in the area. Highest column (Jan 18) corresponds to 18 mm rain.

3. **What are the biogeochemical cycles in urban soil?** Human impact on biogeochemical cycles is a global environmental issue. The pools and fluxes of carbon in urban/suburban soil and its contribution to the global carbon cycle are poorly known. Understanding carbon cycle processes in urban habitats is one of the critical scientific issues recently outlined by an NSF-AGU Committee report [Johnston2003]. Given the enormous heterogeneity of the urban/suburban landscape such assessment is a challenging task. We plan to add $CO_2$ sensors to the motes, and later add other gas sensors (e.g. $CH_4$, $N_2O$). Our measurements will complement data collected at different heights by the Cub Hill carbon flux tower. Ten $CO_2$ rings are currently operating in the Cub Hill area. These rings are sampled monthly. Comparison of different methods will enable us to test the reliability of the sensors in real field conditions.

4. **What is the effect of urbanization on water pathways and what is the coupling of water and carbon storage and flux?** Our pilot study drew a somewhat unexpected interest from hydrologists. As mentioned in the Section 2, soil is an important water reservoir and thus input element in terrestrial hydrology models. Cities have the most heterogeneous landscape due to various land cover and land management. Measurements on soil moisture should reflect this heterogeneity and sensor systems can achieve this goal.

These are ambitious research goals. They would be difficult and expensive to achieve without our current sensor and data analysis infrastructure. But sensor technology is improving rapidly, costs are dropping, and our acquisition and analysis platform is maturing. So, these preliminary research goals will likely expand and be refined as we get more data and experience.

### *5.2 Mote Durability*

To ensure reliable data collection over the long term, both the motes and sensors must perform well under harsh conditions. In our case moisture is the biggest threat. Prior to deployment we performed a 'bathtub test' (enclosures submerged underwater for 24 hours) and a 'freezer test' (enclosures locked in a block of ice and subsequently thawed). These tests caused some boxes to collect moisture (0.3-0.5 ml.) This problem was ameliorated by placing silica desiccant beads in the boxes. The motes operated normally even though they were buried under the snow during the winter (see Figure 10). We found that enclosures became less waterproof after they were opened a few times in the field to update the mote software or replace batteries. In the future, we will avoid opening the boxes by using over-the-air reprogramming, higher capacity batteries, and more aggressive duty cycling. We will pay even more attention to waterproofing the enclosure including the cabling and the antenna holes.

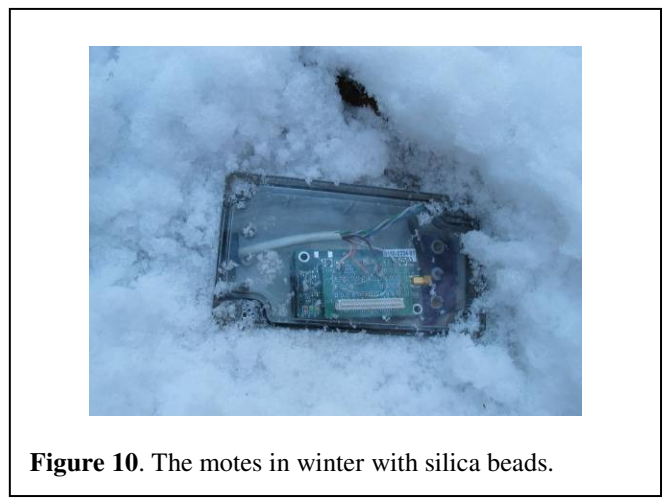

**Figure 10**. The motes in winter with silica beads.

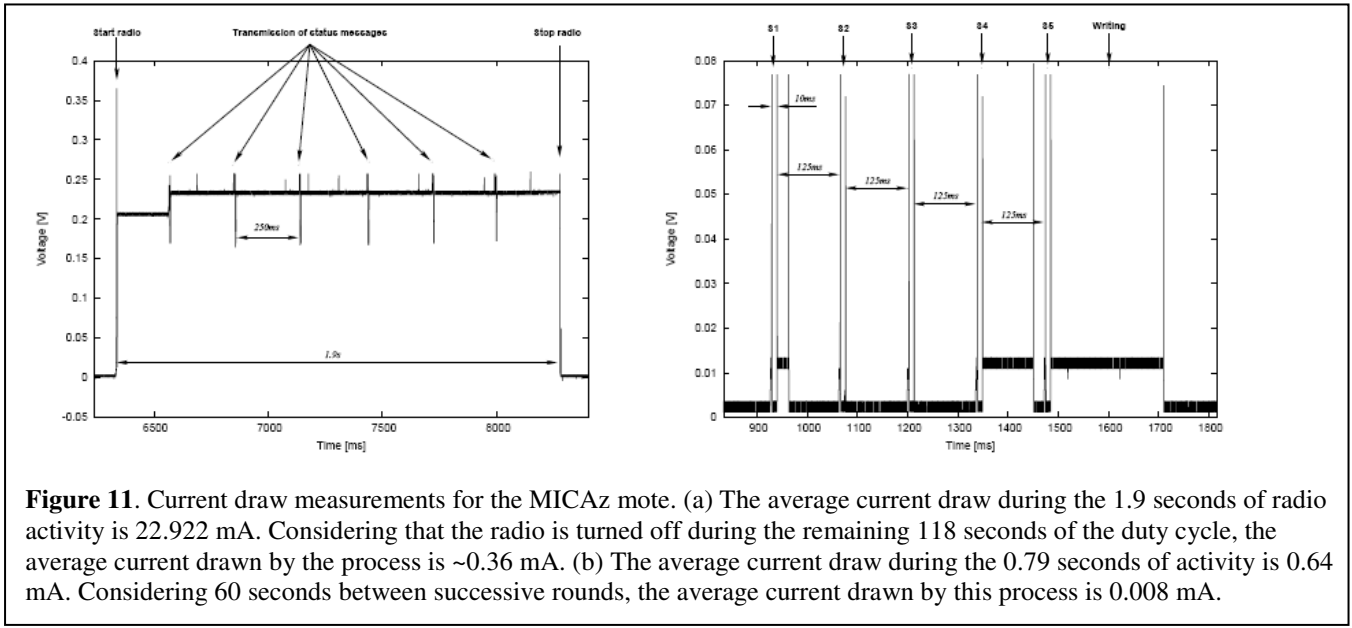

**Figure 11**. Current draw measurements for the MICAz mote. (a) The average current draw during the 1.9 seconds of radio activity is 22.922 mA. Considering that the radio is turned off during the remaining 118 seconds of the duty cycle, the average current drawn by the process is ~0.36 mA. (b) The average current draw during the 0.79 seconds of activity is 0.64 mA. Considering 60 seconds between successive rounds, the average current drawn by this process is 0.008 mA.

Not every sensor worked smoothly and some measurements were lost. For instance mote 60 did not properly record date for about a week in January (see Figure 8). This was probably due to a loose connection. However, we are confident that differences among individual sensors reflect real spatiotemporal heterogeneity. This information should allow soil ecologists to better predict better where and when microbial and invertebrate activity occurs. This activity is tightly coupled with biogeochemical processes such as soil respiration which is an important poorly understood component of the global carbon cycle. Continuous in situ soil monitoring will improve our estimates of soil biota's contributions to these large scale processes.

### 5.3. Mote Energy Consumption

As mentioned before, we power the motes using inexpensive AA Alkaline batteries with an approximate capacity of 2.2Ah. Given the energy budget provided by the batteries, we can derive a first order approximation of mote lifetime by measuring the power consumed by each of the mote's subsystems. We measure the mote's current draw by measuring the voltage differential across a 10Ω resistor placed in series with the device.

Radio is the largest among all energy consumers on the device. Figure (11a) depicts the voltage drop on the resistor during a reporting interval (*i.e.,* when the radio is turned on to send the mote's status reports). This interval lasts 1.9 seconds and 6 status reports are sent in total. Since the radio is turned off during the remaining 118 seconds of the two minute duty cyclel, the average current used by the radio is approximately 0.36 mA[3].

Figure 11b illustrates the power consumed during the sampling of the sensors connected to the mote. During this time the mote samples 5 sensors: soil temperature, soil moisture, enclosure temperature, photo sensor and battery voltage. Each sample is taken by turning the sensor component on, waiting for 10 ms, obtaining a sample from the ADC, and then turning the sensor off. The next sensor is sampled 125 ms later. After all samples have been collected, the mote writes them in its local flash. The whole operation finishes in 0.79 seconds and the average current consumption during this period is 0.64 mA. Since sensor samples are taken every 60 seconds, the overall average current used by the sensors is 0.008 mA. Assuming that current draw is virtually zero when both the radio and the MCU are powered

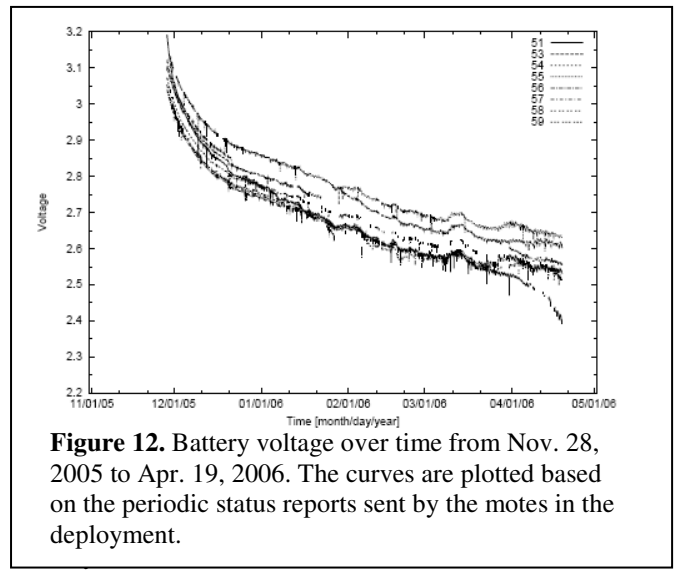

**Figure 12.** Battery voltage over time from Nov. 28, 2005 to Apr. 19, 2006. The curves are plotted based on the periodic status reports sent by the motes in the deployment.

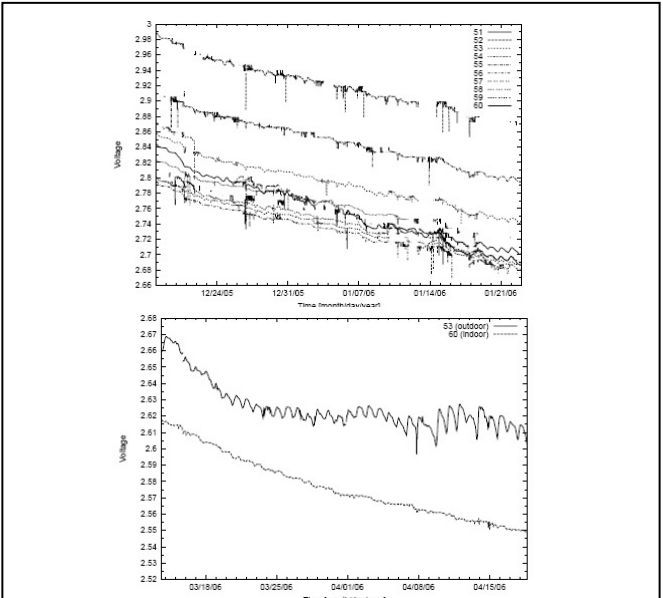

**Figure 13.** Voltage over time for two different periods: winter (left) and spring (right)., Bottom graph shows daily voltage cycles due to outside temperature, while lower curve is voltage of an indoor mote at nearly constant temperature.

off, the average current drawn from the batteries is then 0.368 mA.

Figure 12 illustrates that battery voltage at the deployed motes decreased by approximately 0.4V after 70 days operation (0.2V per battery). Considering that the cutoff voltage of a single battery is 0.8V and using the linear discharge model to approximate remaining battery capacity, as suggested in [Energizer], the measured decrease corresponds to a consumption of 629mAh. This is very close to the 70·24·0.368 = 618mAh consumption computed using the average current drawn by the mote over seventy 24-hour periods. The difference is likely due to the power consumed during data downloads, a factor not included in our analysis. As shown in the following section, the radio is on for an additional one to two minutes during a download transaction. On the other hand, the radio is on for 22 minutes every day to send the periodic status reports. This comparison argues that power consumed during data transfers is not a significant factor when it comes to predicting mote lifetime.

This calculation holds for even smaller scales: the voltage drop during one week is almost 0.02V (cf. Figure 13) corresponding to an expense of 62.8mAh using the linear battery model. For the same period, our average current model estimates energy consumption of 61.8mAh. The high accuracy of this model indicates that it can be used as a planning tool for estimating the lifetime of a mote.

At the same time, there are inherent limitations to this approach, because temperature fluctuations affect the battery voltage. This effect is presented in Figure 13 in which the daily temperature cycle produces noticeable "waves" in voltage readings from motes deployed outdoors, while an identical mote within a building has a smooth and steady discharge slope.

The operating voltage of the MICAz mote, based on its specification, is from 3.6V to 2.7V. However, we determined that motes can reliably operate down to 2.2V before the flash memory stops responding. We also found that the processors and radio operate down to 2.17-2.10V. The current batteries were installed at the end of November. The motes began to stop recording data in mid-April. As of mid-June only three of the ten motes are still reporting. This increased lifetime is a consequence elevated initial voltages in some batteries and a non-linear in battery discharge, illustrated in Figure 13.

## 5.4. Network Quality

Our initial plan was to collect measurements from the motes through a PC, located at the 2$^{nd}$ floor in the Olin building next to our deployment site. The same base station receives the periodic status reports sent by each of the motes. Table 1 presents the number of such reports received over a period of 5 months. Even in the case of motes 52, 58, and 59 that successfully transmitted the most reports, the loss rate was approximately 67%[4]. Since the loss rate was so high, we decided not to use that base station to download collected measurements. Doing so would require excessive retransmissions, which would quickly deplete motes' batteries. On the other hand, even with this high loss rate, periodic reports were used to remotely monitor the network's health [Olin].

**Table 1.** Number of status messages received by the base station from Nov. 28 2005 to Apr. 14 2006. Each mote sent 588,645 reports during this period, which translates to delivery ratios ranging from 29% to 34%.

| MoteID | Reports Received |
|---|---|
| 51 | 189,978 |
| 52 | 197,216 |
| 53 | 188,804 |
| 54 | 182,343 |
| 55 | 187,647 |
| 56 | 187,939 |
| 57 | 191,190 |
| 58 | 197,520 |
| 59 | 197,414 |
| 60 | 168,776 |

---

[4] The base station was offline for small periods of time, but this accounts for at most 3000 packets, less that 0.5% of the total number of packets that should have been received.

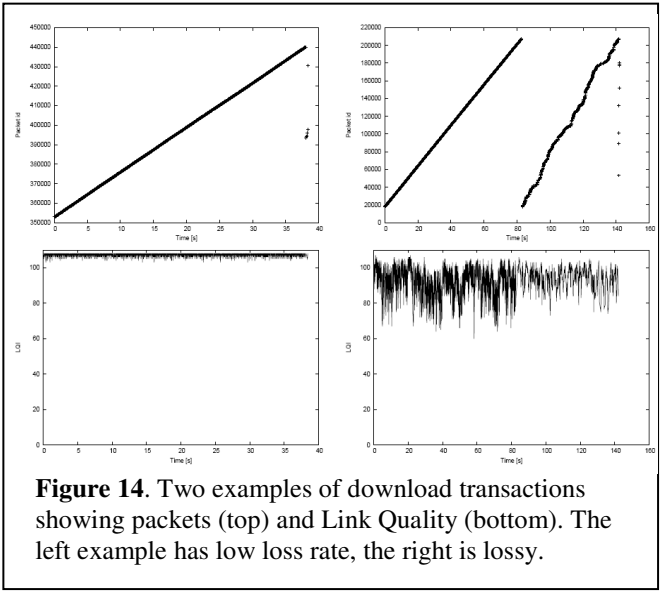

**Figure 14**. Two examples of download transactions showing packets (top) and Link Quality (bottom). The left example has low loss rate, the right is lossy.

The data was collected by 250 downloads between Nov. 28, 2005 and Apr. 20, 2006. A few of them used the fixed base station but most used a laptop located in various positions near the sensor patch to improve signal reception. Figure 14 presents two examples of data downloads performed from the fixed base station: one with low loss rate and another in which more losses were sustained. The top row illustrates the packet identifiers over time while the bottom one shows Link Quality Indicator (LQI) information. During the Bulk Phase of the transfer (Section 2.1.) the good download lost only 6 out of 5438 packets (0.1%) while the bad one lost 689 out of 11811 packets (5.8%). As a result, the *Send-and-Wait* phase during which all the lost packets are retransmitted is more pronounced, with some packets retransmitted twice. Another interesting observation from Figure 14 has to do with the predictive value of received LQI. It is evident that the high quality link has consistently good LQI, while the corresponding LQI of the lossy link displays high variability. As other have proposed, this suggests that LQI could be used to select low-loss links ([Cerpa2005])..

## 6. Discussion and Concluding Remarks

Our primary goal was to demonstrate through an end-to-end data collection prototype that wireless sensor systems could be used in soil monitoring. Even though we did not attempt to meet all the high-level requirements outlined in Section 2.1, building the system proved to be harder than what we expected. While some of our observations have been repeated in the literature (*e.g.,* [Szewczyk2004]), some of them are new.

We learned, as previously reported, that reprogramming is essential for sensor deployments. In our case, we discovered two major software faults after the system was initially deployed. The first bug was related to putting the MCU to sleep mode, while the second one was related to occasional errors when writing to the mote's flash memory. In both cases, we had to retrieve the motes and reprogram them in the lab. Had we used a tool such as Deluge, we would be able to reprogram the motes in the field, decreasing the length of the measurement outage [Deluge].

Contrary to the hype, sensor motes are still expensive. We estimated the cost per mote including the main unit, sensor board, custom sensors, enclosure, and the time required to implement, debug and maintain the software to be around $1,000, equivalent to the price of a mid-range PC! Calibrating each of the sensors costs more than the sensors themselves -- and is not a task for novices. While equipment costs will eventually be reduced through economies of scale, there is clearly a need for standardized connectors for external sensors and in general a need to minimize the amount of custom hardware necessary to deploy a sensor system. Unfortunately, sensor and mote vendors seem to want proprietary interfaces that limit our ability to add 3rd party sensors. We hope future motes use standard connectors.

We also found that low-level programming is a necessary and challenging task when building sensor systems for new applications. Not only did we have to write low-level device drivers for the soil temperature and moisture sensors, but also for power control, as well as for calibration procedures. Moreover, using acquisitional processors such as TinyDB [Madden2003] was not an option in our case given the requirement to collect *all* the data.

Finally, we identified a need for system design and deployment tools that instruct scientists where to place gateways and sensor relay points that can help transport collected measurements back to an Internet-connected base station [Burns2006]. These tools will replace the current trial-and-error, labor-intensive process of manual topology adjustments that disturbs the deployment area.

A wireless sensor network is only the first component in an *end-to-end* system that transforms raw measurements to scientifically significant data and results. This end-to-end system includes calibration, interfaces with external data sources (*e.g.,* weather data), databases, Web Services interfaces, analysis, and visualization tools.

While the sensor network community has focused its attention on routing algorithms, self-organization, and in-network processing among other things, environmental monitoring applications[5] require a different emphasis: reliable delivery of the majority (if not all) of the data and metadata, high quality measurements, data storage, analysis, and visualization, and reliable operation over long deployment cycles. We believe that focusing on these problems will lead to interesting new avenues in sensor network research.

---

[5] Sometimes derided as *academically dull applications*, a characterization with which the ecologist in our team does not agree.

## Acknowledgments

We would like to thank the Microsoft Corporation, the Seaver Foundation, and the Gordon and Betty Moore Foundation for their support. Răzvan Musăloiu-E. is supported through a partnership fund from the JHU Applied Physics Lab. Josh Cogan is partially funded through the JHU Provost's Undergraduate Research Fund. Andreas Terzis is partially supported by NSF CAREER grant CNS-0546648.